\documentclass[12pt]{article}
\usepackage{siunitx}
\usepackage{graphicx}
\usepackage[utf8]{inputenc}
\usepackage{graphicx,xcolor,setspace,geometry}
\usepackage{amsmath,setspace}
\usepackage{amssymb}
\usepackage{mathtools}
\usepackage{multirow}
\usepackage{booktabs}
\usepackage{threeparttable}
\usepackage{adjustbox}
\usepackage{nicefrac}
\usepackage{natbib}
\usepackage{multirow}
\usepackage{color}
\usepackage{tikz}
\usetikzlibrary{shapes}
\usetikzlibrary{plotmarks}
\usetikzlibrary{tikzmark,matrix,calc}
\usetikzlibrary{arrows.meta,calc,decorations.markings,math,arrows.meta}
\usepackage{caption}
\usepackage{hyperref}

\title{Markov-modulated marked Poisson processes \\ for modelling disease dynamics based on \\ medical claims data}

\author{Sina Mews$^{1,}\footnote{Corresponding author; email: \texttt{sina.mews@uni-bielefeld.de}.}$, Bastian Surmann$^{2}$, Lena Hasemann$^{2}$, \\ and Svenja Elkenkamp$^{2}$ \\
 \\
$^{1}$Department of Business Administration and Economics, \\ Bielefeld University, Germany \\
$^{2}$Department for Health Economics and Health Care Management, \\ Bielefeld University, Germany}

\date{}

\begin{document}

\begin{spacing}{1.25}
    \maketitle
\end{spacing}

%\vspace{-5mm}

\begin{spacing}{1.5}

\begin{abstract}
We explore Markov-modulated marked Poisson processes (MMMPPs) as a natural framework for modelling patients' disease dynamics over time based on medical claims data. 
In claims data, observations do not only occur at random points in time but are also informative, i.e.\ driven by unobserved disease levels, as poor health conditions usually lead to more frequent interactions with the healthcare system.
Therefore, we model the observation process as a Markov-modulated Poisson process, where the rate of healthcare interactions is governed by a continuous-time Markov chain. 
Its states serve as proxies for the patients' latent disease levels and further determine the distribution of additional data collected at each observation time, the so-called marks.
Overall, MMMPPs jointly model observations and their informative time points by comprising two state-dependent processes: the observation process (corresponding to the event times) and the mark process (corresponding to event-specific information), which both depend on the underlying states. 
The approach is illustrated using claims data from patients diagnosed with chronic obstructive pulmonary disease (COPD) by modelling their drug use and the interval lengths between consecutive physician consultations.
The results indicate that MMMPPs are able to detect distinct patterns of healthcare utilisation related to disease processes and reveal inter-individual differences in the state-switching dynamics.
\end{abstract}


\section{Introduction}
\label{s:intro}

Whenever a patient interacts with the healthcare system, claims data are routinely collected from all providers caring for the patient (such as physicians, pharmacies, or hospitals), containing information on costs, patient-specific diagnoses, and received treatments like medication.
These rich databases on real-life healthcare provisions are increasingly prominent in public health research as well as decision-making processes of different stakeholders. 
For example, claims data are used to optimise health service provisions \citep[e.g.][]{dutta2022}, estimate the prevalence and incidence of diseases \citep[e.g.][]{nerius2017}, or identify patients at risk of hospital readmission \citep[e.g.][]{min2019}. 
While claims data have been extensively used for disease prediction \citep[see, e.g.,][]{nielsen2017, christensen2018, hossain2019}, much less attention has been paid to their vast amount of (implicit) information on disease dynamics over time, such as (changes in) medication, hospital stays, or the frequency of physician consultations (but see \citealp{ploner2020}).
In this contribution, we thus draw on these comprehensive data sets to model the temporal courses of diseases and to learn about patients’ health conditions over time.
Gaining insights into the latter is crucial to assess treatment effects on disease progression and to distinguish health conditions associated with different demands of care. 
These results, in turn, can be used to evaluate individual as well as economic consequences of specific diseases. 

For modelling patients' disease processes over time, medical claims data pose two main challenges, namely
(1) that patients' disease stages are not (directly) observed and
(2) that patients' interactions with the healthcare system are driven by the disease stages themselves, with more frequent interactions when the patient's condition is poor. 
Regarding (1), while claims data provide extensive records on patients’ received services and diagnoses, they lack information on the actual health condition, specifically on distinct disease stages (such as mild, moderate, and severe) graded according to disease-specific practice guidelines. 
Therefore, the unobserved evolution of patients' disease activity over time needs to be inferred from available medical data. 
A popular approach for estimating properties of the disease process underlying the observed data are (continuous-time) hidden Markov models (HMMs; e.g.\ \citealp{bureau2003, jackson2003}). 
Although these latent-state models have been successfully applied in different studies on disease progression, for example based on monthly MRI scans \citep{altman2005} or screening data \citep{amoros2019}, they are not appropriate for modelling routinely collected claims data due to the second (2) challenge: 
whereas (continuous-time) HMMs assume that observation times are \textit{non-informative} (i.e.\ independent of the underlying disease process), observation times of claims data are \textit{informative} (i.e. dependent on a patient's disease activity), as healthcare interactions are predominantly initiated on demand by the patient. 
Specifically, a patient's poor health condition usually leads to more frequent and hence clustered interactions over time --- a key characteristic of claims data that HMMs cannot account for. 

A natural approach to model the occurrence and clustering of events over a continuous time interval is provided by Markov-modulated Poisson processes (MMPPs). 
MMPPs generalise homogeneous Poisson point processes in that they assume the event rate to depend on a latent finite-state Markov process in continuous time. 
According to this underlying discrete-valued Markov process, the event rates switch between different levels over time, thus resulting in clustering of events. 
These events can, for example, correspond to mouse clicks (i.e.\ Web page requests; \citealp{scott2003}), the surfacing of whales \citep{langrock2013}, or the occurrence of earthquakes \citep{lu2012}.  
While MMPPs have been applied in various areas, applications in the medical context and especially to disease modelling are rare. 
However, \citet{lange2015} use MMPPs to jointly model patients' visit times and their observed (though possibly misclassified) disease process based on electronic health records (EHR) data.
Although such information on patients' health condition is unavailable in claims data, it is reasonable to assume that an increased disease activity results in higher healthcare utilisation \citep[see, e.g.,][]{lange2015, alaa2017, gasparini2020, su2021}. 
Consequently, claims data contain information on the underlying disease process by accurately reflecting (the frequency of) patients' interactions with the healthcare system, which often occur in clusters (see Figure~\ref{fig:exTrajectory}).
These healthcare interactions can thus be modelled as an MMPP, where the rate of interactions is governed by a continuous-time Markov chain, whose states serve as a proxy for the patients' unobserved disease levels.

At each interaction time with the healthcare system, claims data provide additional observations that can help to improve inference on the underlying disease process.
To infer a patient's disease level using both the clustered interaction times and additional data observed at these interactions, we propose to analyse patients’ disease activity over time using Markov-modulated \textit{marked} Poisson processes (MMMPPs).
MM\textit{M}PPs extend MMPPs by jointly modelling the observation times as well as data collected at these observation times, the so-called \textit{marks}.
These marks --- just like the event rates --- depend on the underlying discrete-valued Markov process, whose latent states determine the distribution of the marks. 
In summary, MMMPPs thus consist of three stochastic processes: the state process, the observation process, and the mark process. 
The state process, corresponding to the unobserved Markov process, governs both the observation process, consisting of the observation times, and the mark process, consisting of the data collected at each observation time (i.e.\ the marks). 
The latter may correspond to treatments such as (the amount of) drug use or costs associated with each healthcare interaction. 
To our knowledge, only \citet{alaa2017} use a similar modelling approach in the medical context, namely the semi-Markov-modulated marked Hawkes process, to model hospitalised patients’ latent clinical states over time. 
The main difference to our approach is that \citet{alaa2017} focus on risk prognosis based on EHR data, particularly patients’ vital signs and lab tests, whereas we aim to extract information on patients’ health conditions over time from claims data.
While claims data pose particular challenges as outlined above, these can be addressed naturally by MMMPPs, which allow inference on the latent disease process based on observations occurring at informative and clustered points in time.

\section{Markov-modulated marked Poisson processes}
\label{s:methods}

\subsection{Basic model formulation}
\label{ss:basicModel}

We consider (claims) data containing information on the random observation times $T_0,$ $T_1,\ldots,T_n$, $0=T_0<T_1<\ldots<T_n$, which occur at irregularly spaced points in time, as well as additional data $Y_{t_1},\ldots,Y_{t_n}$ collected at the realised observation times.
These sequences of random variables are referred to as the observation process and the mark process, respectively, and depend on an underlying, unobserved state process $\{S_t\}_{t \geq 0}$. 
From now on, let the integer $\tau = 1, 2, \ldots, n$ denote the index of the observation in the sequence, such that $Y_{t_\tau}$ and $S_{t_\tau}$ shorten to $Y_\tau$ and $S_\tau$, respectively.

The state process is modelled as an $N$-state continuous-time Markov chain.
Transitions between the states are governed by a transition intensity matrix $\mathbf{Q} = (q_{ij})_{i,j = 1, \ldots, N}$, whose off-diagonal elements $q_{ij} \geq 0$, $i,j=1,\ldots,N$, $i \neq j$, can be interpreted as the rates at which transitions from state $i$ to state $j$ occur.
The duration in each state $i = 1, \ldots, N$ is exponentially distributed with parameter $ q_{ii} = \sum_{j \neq i} q_{ij} $, where $-q_{ii}$ is the $i$-th diagonal entry in $\mathbf{Q}$.
Furthermore, the initial distribution of the state process is denoted by $\boldsymbol{\delta} = (\delta_1, \ldots, \delta_N),$ where $\delta_i = \text{Pr}(S_0 = i)$. 
In the context of claims data, the states represent disease levels associated with different frequencies of healthcare interactions (as reflected in the observation process) and different distributions of variables collected at each observation time (as reflected in the mark process).
It is important to note that although the states can approximate distinct disease levels, they are purely data-driven, as the model just picks up the strongest patterns in the data. 
With no information on patients' actual health condition available, the states thus do not (necessarily) correspond to any formally defined disease stages based on practice guidelines or grading systems. 

The observation process is modelled as a doubly stochastic point process, namely a Markov-modulated Poisson process (MMPP) whose event rates $\lambda_i, i=1, \ldots, N,$ are selected by the underlying Markov chain.
Conditional on the current state $i = 1, \ldots, N$ being active throughout the time interval [0, $t$], the number of events in [0, $t$] follows a Poisson distribution with parameter $\lambda_i t$.
Furthermore, the waiting times between consecutive events $X_\tau = T_\tau - T_{\tau - 1}, \tau = 1, \ldots, n$, are exponentially distributed with parameter $\lambda_i$ within each state. 
In the context of modelling medical claims, the frequency, or rate, of observations (i.e.\ healthcare interactions) $\lambda_i$ is thus determined by the underlying (disease) state $i = 1, \ldots, N$.
We assume that severe disease levels are associated with higher event rates $\lambda_i$, as an increased disease activity (usually) causes higher healthcare utilisation. 
Specifying the observation process as an MMPP thus accounts for the time-varying intensity of the observations and their temporal dependence.

For the mark process, we assume the distribution of a variable (i.e.\ mark) $Y_\tau$ collected at observation time $t_\tau$ to be fully determined by its underlying state. 
In particular, the conditional independence assumption $ f(y_\tau | t_1, \ldots, t_\tau, y_1, \ldots, y_{\tau - 1}, s_1, \ldots s_\tau) = f(y_\tau | s_\tau) $ implies that the underlying Markov chain selects which state-dependent distribution $f_i(y_\tau) = f(y_\tau | s_\tau = i)$ is active at time $t_\tau$.
As the state-dependent distributions can take on any (parametric) form, various data types, such as binary, count, or continuous variables, can be considered in the mark process. 
Note that while the marks are conditionally independent of all previous (and future) marks, observation times, and states, the state process induces correlation in the mark (as well as the observation) process.
For now, we assume all stochastic processes to be homogeneous, but will discuss how to relax this assumption in Section~\ref{ss:covs}.

\subsection{Likelihood evaluation and maximisation} 

Subject to the unobserved Markovian state process, the MMMPP jointly models the mark process and the observation process. 
To evaluate the corresponding likelihood of the model, inferential tools from the HMM framework, in particular the corresponding efficient algorithms for parameter estimation, can be applied \citep{lu2012}.
In contrast to \citet{lu2012}, who estimates the model parameters based on the EM algorithm, we numerically maximise the likelihood using the HMM-based forward algorithm \citep[see, e.g.,][]{zucchini2016}. 
Defining the observation process by its waiting times, the likelihood of the observed sequence $\{ (x_{\tau}, y_{\tau}) \}_{\tau \in \{1, \ldots, n\} }$ is given by:
\begin{equation}
    \mathcal{L} = \boldsymbol{\delta} \mathbf{P}(y_0) \Bigl( \prod_{\tau=1}^n \text{exp}\bigl( (\mathbf{Q} - \boldsymbol{\Lambda}) x_\tau \bigr) \boldsymbol{\Lambda} \mathbf{P}(y_\tau) \Bigr) \boldsymbol{1},
\label{e:llk}
\end{equation}
where $\mathbf{P}(y_\tau) = \text{diag}\bigl\{ \Pr(y_\tau | s_\tau=1), \ldots , \Pr(y_\tau | s_\tau=N) \bigr\}$ and $\boldsymbol{\Lambda} = \text{diag}\bigl\{ \lambda_1, \ldots , \lambda_N \bigr\}$ are diagonal matrices and $\boldsymbol{1} \in \mathbb{R}^N$ denotes a column vector of ones \citep{lu2012}.
Note that different methods to compute the matrix exponential $\text{exp}(\mathbf{A}) = \sum_{d=0}^\infty \mathbf{A}^d / d!$ can be used \citep{moler2003}. 
Equation~(\ref{e:llk}) corresponds to a recursive calculation of the likelihood, where the joint likelihood of the observed sequence up to time $t_{\tau}$ is updated based on the likelihood up to time $t_{\tau-1}$, retaining information on the probabilities of the different states being active. 
To avoid numerical under- or overflow in the maximisation, a scaling strategy needs to be applied (see \citealp{zucchini2016} for further details on technical issues arising in the likelihood maximisation).

\subsection{Incorporating covariates}
\label{ss:covs}

Covariates can be incorporated in all three components of the MMMPP using a general linear regression framework. 
Depending on whether the covariates affect the switching dynamics of the state process, the event frequency of the observation process, or the state-dependent distributions of the mark process, the respective parameters can be modelled as a function of both individual-specific and time-varying covariates.

Regarding the mark process, for example the mean of the state-dependent distributions can be specified as $\mu_i^{(t)} = g^{-1}(\beta_0^{(i)} + \beta_1^{(i)} z_t)$, where $z_t$ is a time-varying covariate, $\beta^{(i)}$ are state-dependent coefficients, and $g^{-1}()$ denotes an inverse link function accounting for possible parameter constraints. 
In contrast to the mark process, modelling the state process or the observation process as a function of covariates is straightforward only if covariate values are constant across time. 
For instance, given a time-constant, individual-specific covariate $z^{(k)}$ for the $k$-th person in the data set, the (off-diagonal) state transition intensities can be specified as $q_{ij}^{(k)} = \text{exp}(\beta_0^{(ij)} + \beta_1^{(ij)} z^{(k)})$, $i \neq j,$ and the event rates as $\lambda_i^{(k)} = \text{exp}(\beta_0^{(i)} + \beta_1^{(i)} z^{(k)}) $, respectively.
In this case, the parameters of the respective process differ between individuals, but importantly, they remain constant over time.
If instead, a covariate varies (continuously) across time, leading to (either) nonhomogeneous state transition intensities or event rates, the resulting matrix exponentials in Equation~(\ref{e:llk}) are not analytically tractable. 
However, the latter can be calculated recursively over shorter intervals with constant parameter values.
For time-varying covariates that are piecewise constant over time --- such as the age of a person measured in years --- the matrix exponentials are thus calculated separately for each time interval on which the covariate, and hence the parameters $q_{ij}^{(t)}$ or $\lambda_i^{(t)}$, respectively, are constant \citep[see, e.g.,][]{fouchet2016, choquet2018}. 
Similarly, for covariates varying continuously over time --- such as the time since initial diagnosis --- their effects and hence, the respective parameters $q_{ij}^{(t)}$ or $\lambda_i^{(t)}$ can be approximated by (constant) step functions (see, e.g., \citealp{langrock2013} or \citealp{mews2022} for a detailed account of the approach).

\section{Case study: Modelling disease activity of COPD patients} 
\label{s:results}

\subsection{Data description and model formulation}

We consider data from one of the largest statutory health insurance (SHI) companies in Germany, providing comprehensive information on interactions of the insured with the healthcare system. 
Covering a period of 16 years, namely from 2005 to 2020, the data comprise, inter alia, the insured persons' sex and age as well as their diagnoses (based on ICD-10 classification), prescriptions, hospitalisations, sick days, and associated healthcare costs.
Consequently, the data set allows not only to identify particular groups of patients suffering from a certain disease but also to derive details on their health condition.
In this case study, we focus on patients diagnosed with chronic obstructive pulmonary disease (COPD), which was previously analysed in similar studies on latent-state modelling \citep{luo2021}. 
COPD is a common lung disease with a prevalence of 6.37\% in Germany in 2017 \citep{akmatov2019}. 
It is associated with persistent respiratory symptoms and can be treated to slow, though not fully reverse, its progression.
Due to its (usually) slow progression, longitudinal data are necessary to gain insight into the disease process of COPD over time.

To cover the longest possible observation period and to reduce heterogeneity in our study population, we consider persons initially diagnosed with COPD in 2008 (using the years 2005--2007 as a pre-observation period; see Appendix~\ref{app1} for detailed information on the criteria used to select the study population).  
As COPD is often associated with various comorbidities, which hinder inference on the disease dynamics, we calculate the age-adjusted Charlson comorbidity index (ACCI) based on \citet{quan2011} for each person.
Data from persons with an ACCI maximum score larger than four (corresponding to severe age-adjusted comorbidities; \citealp{bannay2016}) were excluded from analyses.
As a consequence, our study population is younger than COPD patients in the general population. 
The final data set includes 470 persons (141 males and 329 females) with a median age of 49 years (min: 21; max: 69) at initial diagnosis. 
Regarding the ACCI, most persons (80\%) show a moderate age-adjusted comorbidity burden (i.e.\ an ACCI of 3 or 4), while the rest (20\%) have mild age-adjusted comorbidities (i.e.\ an ACCI of 1 or 2; \citealp{bannay2016}. 
Overall, the final data comprise 112,297 observations, covering a mean observation period of 12 years per person (min: 6; max: 13) after initial COPD diagnosis. 

\begin{figure}[t]
    \centering
    \includegraphics[width = 0.85\textwidth]{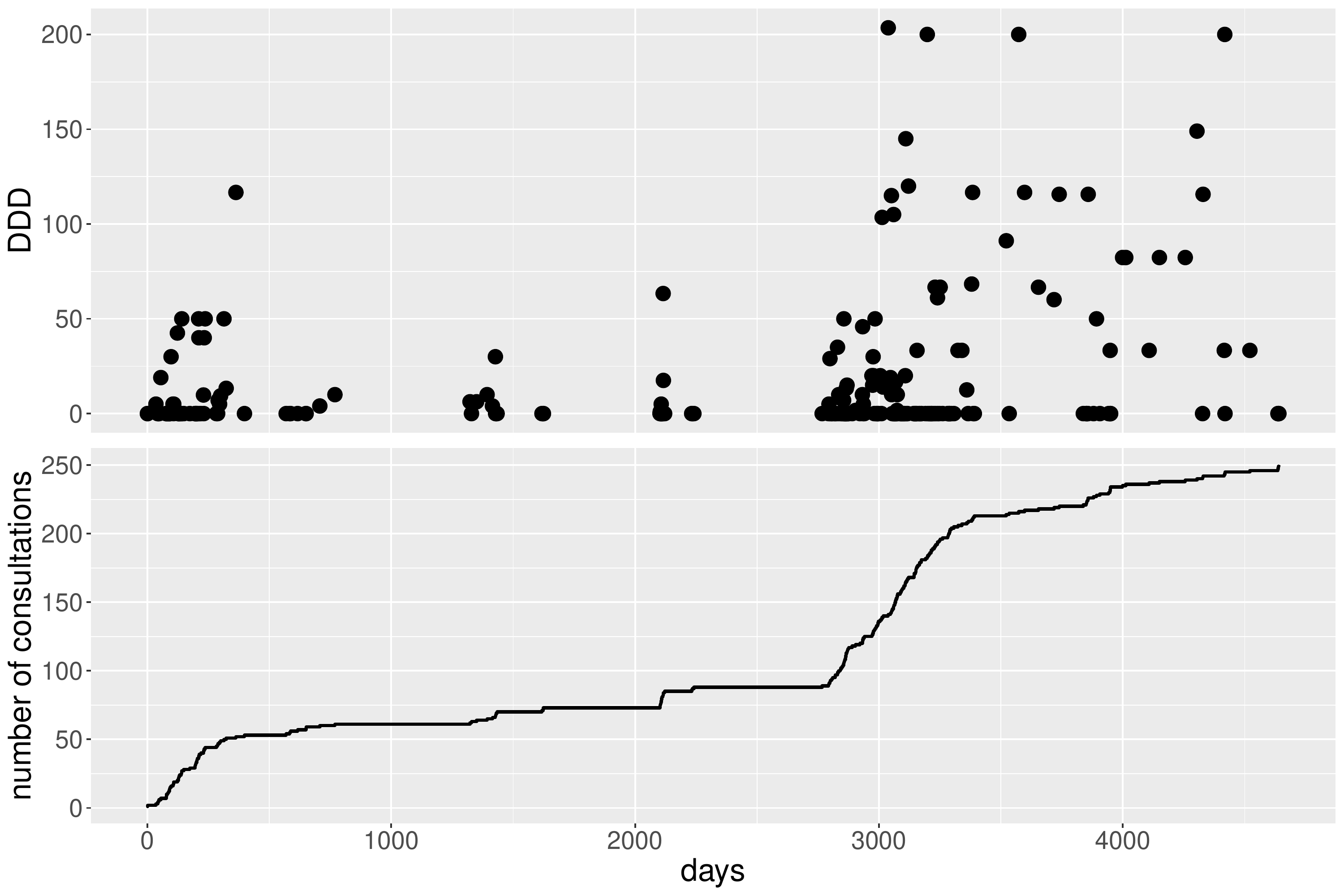}
    \caption{Example DDD sequence (upper plot) and step function of the number of consultations over time (lower plot) for one patient. 
    }
    \label{fig:exTrajectory}
\end{figure}

For modelling COPD patients' disease activity over time, we consider both the interval length between consecutive physician consultations (i.e.\ the waiting times) and patients' drug use.
The latter is measured in daily defined doses (DDDs) --- a standardised unit for drug consumption --- based on physicians' prescriptions contained in the SHI data.
As not every consultation is associated with a drug prescription, we have a large amount of zero DDDs (i.e.\ no prescribed drugs; 59.5\%) in our data.
If the patient received a prescription, the average amount of DDDs prescribed is 106 (sd: 115). 
Regarding the waiting times, on average 18 days pass between consecutive physician consultations (sd: 25). 
Histograms indicating the distribution of the observed variables are shown in Figure~\ref{fig:histDens} in the Appendix. 
Furthermore, an example trajectory of a patient's DDDs over time together with the accumulative number of consultations is shown in Figure ~\ref{fig:exTrajectory}, reflecting the clustered structure of the observations.
In particular, one can easily distinguish periods with high (e.g.\ 12 consultations within 31 days) and low healthcare utilisation (e.g.\ more than one year between consecutive consultations). 
The relation between the frequency of physician consultations and DDDs, in contrast, is not as obvious:
there is a tendency to observe more consultations without drug prescription (i.e.\ zero DDDs) if the frequency of interactions is high, but no clear patterns regarding the amount of DDDs prescribed is evident. 

\subsection{Model formulation}

We model the observed waiting times between consecutive physician consultations and the DDDs prescribed as an MMMPP, where the underlying states can be interpreted as patients' health condition over time. 
Regarding the observation process, this means the waiting times are modelled using state-dependent exponential distributions, while for the mark process, we assume the DDDs to follow a zero-adjusted gamma distribution with state-specific parameters. 
As we are interested in inter-individual differences in the state-switching dynamics, we model the state transition intensities as a function of patients' sex, their age at initial diagnosis (\textit{ageD}), and the ACCI (dichotomised into either mild or moderate age-adjusted comorbidities):
\begin{equation*}
    q_{ij} = \text{exp}\left(\beta_0^{(ij)} + \beta_1^{(ij)} \textit{sex} + \beta_2^{(ij)} \textit{ACCI} + \beta_3^{(ij)} \textit{ageD} \right), \quad \text{for} \hspace{1mm} i \neq j.
\end{equation*}
For simplicity, we restrict ourselves to a 2-state model, noting that the methodology is generally applicable for any finite number of states. 
Arguably, this is only one of many possible model formulations that can be considered (cf.\ Section~\ref{s:discussion}) --- however, presenting a comprehensive analysis of COPD patients' disease progression is beyond the scope of this case study. 
The parameters of the MMMPP, namely the regression coefficients and the state-dependent rate parameters as well as the parameters of the gamma distributions, are estimated by numerically maximising the joint likelihood over all patients, which is calculated as the product of the individual likelihoods given in Equation~\ref{e:llk}.
% The R code for model fitting (and subsequent state decoding) is available in Online Supplementary Material 2. 

\subsection{Results}

The parameter estimates of both the observation and the mark process are presented in Table~\ref{tab:estParams}, while plots visualising the estimated state-dependent distributions are shown in Figure~\ref{fig:histDens} in the Appendix. 
Regarding the observation process, the rate of healthcare interactions in state 1 is roughly $\nicefrac{1}{8}$, whereas in state 2, it is $\nicefrac{1}{34}$.
Therefore, we would expect one consultation every 8 days in state 1 compared to one consultation every 34 days in state 2.
Regarding the mark process, while the estimated gamma distribution in state 1 comprises, on average, lower DDDs than the one in state 2, the estimated variances in both states are high. 
Consequently, there is substantial overlap of both estimated gamma distributions. 
In state 1, however, the probability of observing no prescription (i.e.\ a zero DDD) is much higher than in state 2, namely by 36 percentage points.
The latter appears reasonable: if a patient has many physician consultations in a short time interval (e.g.\ because of different examinations or tests), they will not receive a prescription at each of these visits. 
Taking into account the state-dependent interaction rates, the expected sum of DDDs per month (30 days) is 245 in state 1 compared to 208 in state 2, indicating that although the amount of prescribed drugs at a single consultation is expected to be lower in state 1, the higher interaction rate leads to a larger expected number of DDDs than in state 2. 
Overall, state 1 is thus characterized by frequent healthcare interactions and higher drug use (over a longer time period), which could be interpreted as a state of poor health condition or, alternatively, a period in which a treatment needs to be appropriately adjusted to a patient (e.g.\ right after disease diagnosis) --- an interpretation supported by the estimated initial state probabilities, as it is 2.5 times more likely that a person is in state 1 rather than state 2 right after their initial COPD diagnosis.
In contrast, state 2 consists, on average, of approximately monthly healthcare interactions with higher DDDs at a single consultation and hence could be interpreted as a state in which a patient's therapy is maintained.
Therefore, we (tentatively) describe state 1 as the high and state 2 as the low disease level. 
Importantly, however, these disease states are derived in a data-driven way and as such should not be expected to match disease stages postulated in the literature exactly.
 
\begin{table}[t]
    \centering
    \caption{(State-dependent) parameter estimates with 95\% confidence intervals (CIs) for the initial state distribution, the observation process and the mark process, where the (zero-adjusted) gamma distribution is parameterised in terms of its mean and standard deviation. The CIs were calculated based on the observed Fisher information.}
    \label{tab:estParams}
    \resizebox{\columnwidth}{!}{%
    \begin{tabular}{llccccc} % @{\hskip 0.5in}
    \\[-1.8ex]\hline 
\hline \\[-1.8ex] 
        \multicolumn{2}{l}{\multirow{2}{*}{parameter}} & \multicolumn{2}{c}{estimate} & & \multicolumn{2}{c}{95\% CI} \\ \cline{3-4} \cline{6-7}
        & & state 1 & state 2 & & state 1 & state 2 \\ \hline
        $\delta_i$ & (initial state prob.) & 0.707 & 0.293 & & [0.633;0.771] & [0.229;0.367] \\
        $\lambda_i$ & (rate of exponential dist.) & 0.123 & 0.029 & & [0.121;0.125] & [0.029;0.030] \\
        $\mu_i$ & (mean of gamma dist.) & 80.9 & 124.5 & & [79.1;82.7] & [122.8;126.3] \\
        $\sigma_i$ & (std.\ deviation of gamma dist.) & 85.0 & 116.9 & & [82.8;87.2] & [115.0;118.8] \\
        $\pi_0^{(i)}$ & (prob.\ at zero) & 0.726 & 0.370 & & [0.721;0.730] & [0.362;0.379] \\ \hline
    \end{tabular}
    }
\end{table}

\begin{table}[t]
    \centering
    \caption{Expected duration (in days) as well as expected percentage of time spent in each state with 95\% CIs, calculated for different groups (based on the estimated regression coefficients presented in Table~\ref{tab:coefSI} in the Appendix). CIs were obtained based on Monte Carlo simulation from the estimators’ approximate distribution as implied by maximum likelihood theory. The reference group are female patients with moderate comorbidities and age 49 at initial diagnosis (corresponding to the mean age at diagnosis in the study population). For all other groups, only the specified characteristic is changed with regard to the reference group (for instance, the second row are male patients with moderate ACCI and mean age at diagnosis).}
    \label{tab:effects}
            \begin{tabular}{lcccc} % @{\hskip 0.5in}
        \\[-1.8ex]\hline 
\hline \\[-1.8ex] 
        \multirow{2}{*}{group} & \multicolumn{4}{c}{expected duration (in days)} \\ \cline{2-5} 
        & \multicolumn{2}{c}{state 1} & \multicolumn{2}{c}{state 2} \\ \hline
        reference group & 48.2 & [45; 51.7] & 94.2 & [88.3; 100.3]  \\
        male & 49.1 & [44.7; 54.3] & 175.1 & [158.4; 194.4]  \\
        mild ACCI & 53.5 & [47.3; 60.6] & 148.3 & [131.7; 167.8]  \\
        min.\ age: 21 & 61.3 & [52.0; 72.7] & 77.0 & [65.9; 90.1]  \\
        max.\ age: 69 & 40.7 & [36.2; 45.7] & 108.4 & [96.9; 121.4]  \\ \hline
        \multirow{2}{*}{group} & \multicolumn{4}{c}{expected percentage of time} \\ \cline{2-5}
        & \multicolumn{2}{c}{state 1} & \multicolumn{2}{c}{state 2} \\ \hline
        reference group & 33.8\% & [32.8; 35.0] & 66.2\% & [65.0; 67.2]  \\
        male & 21.9\% & [20.7; 23.2] & 78.1\% & [76.8; 79.3]  \\
        mild ACCI & 26.5\% & [24.9; 28.3] & 73.5\% & [71.7; 75.1]  \\
        min.\ age: & 44.3\% & [41.5; 47.2] & 55.7\% & [52.8; 58.5]  \\
        max.\ age: & 27.3\% & [25.7; 28.8] & 72.7\% & [71.2; 74.3]  \\ \hline
    \end{tabular}
\end{table}

The estimated regression coefficients governing the state process are presented in Table~\ref{tab:coefSI} in the Appendix.
From these, we can calculate the transition intensity matrix $\boldsymbol{Q}$ for different covariate values, from which in turn we can derive the expected durations within each state (cf.\ Section~\ref{ss:basicModel}). 
Results are presented in Table~\ref{tab:effects}.
The reference group of female patients with moderate comorbidities and age 49 at initial diagnosis (corresponding to the mean age at diagnosis in the study population) is expected to spend roughly 48 days in state 1 compared to 94 days in state 2.
Thus, we would expect a person from the reference group to spend approximately one third of her time in the high disease level and two thirds in the low disease level (corresponding to the stationary distribution of the latent Markov chain given the respective covariate values).
In contrast, male patients with otherwise the same characteristics spend considerably more time in the low disease level (i.e.\ 11.9 percentage points more), mainly because they remain in state 2 for a longer time period before switching to state 1. 
A similar pattern is found for mild (in contrast to moderate) comorbidities, with a difference of 7.3 percentage points in the expected time spent in each state.
Regarding age at diagnosis, there is a clear effect that younger persons spend both more time in state 1 and less time in state 2 compared to the reference group, which is reversed in older age (cf.\ Figure~\ref{fig:ageEffect} in the Appendix, which plots the expected percentage of time spent in the high disease level for the whole age range observed in the study population). 

\begin{figure}[t]
    \centering
    \includegraphics[width = 0.85\textwidth]{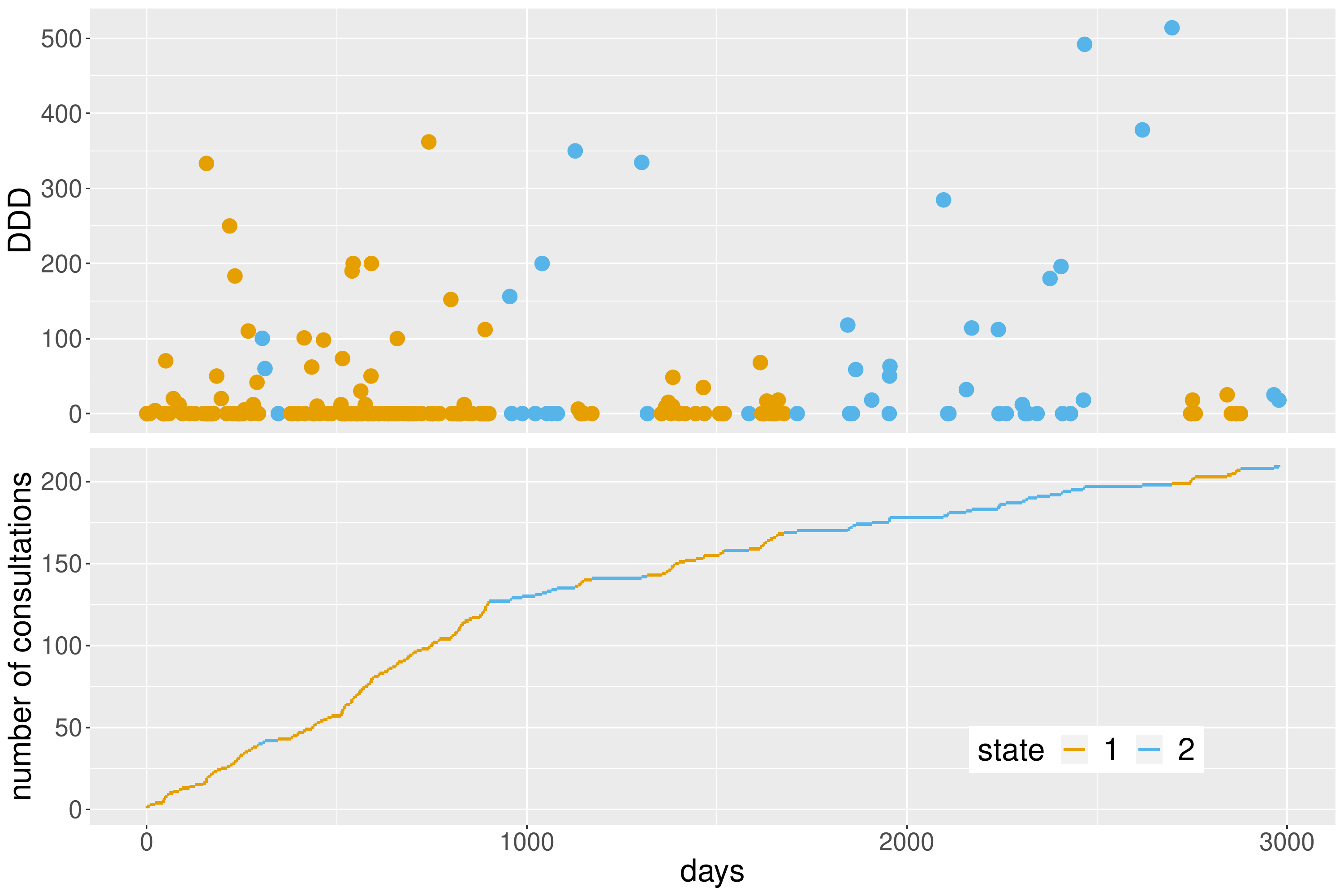}
    \caption{Example DDD sequence (upper plot) and step function of the number of consultations over time (lower plot) for one patient, coloured according to the decoded states at each observation time.} 
    \label{fig:decStates}
\end{figure}

Based on the fitted MMMPP, we can use the Viterbi algorithm \citep{viterbi1967, forney1973} to infer the most probable sequence of latent (disease) states for each person.
These decoded state sequences provide insight into the individual course of a disease by classifying each observation into either the high or low disease level (see Figure~\ref{fig:decStates}). 
Overall, the model appears to adequately distinguish (qualitatively) different periods of healthcare utilisation and drug use, while taking into account the temporal dependence structure of the data.
To further assess the model fit, we checked if our MMMPP could reproduce the patterns found in the SHI data by simulating observations from the estimated model.  
The resulting synthetic data reflect the distributions of both the waiting times between consecutive observations and the DDDs quite well (cf.\ Figure~\ref{fig:simDist} in the Appendix).
However, regarding the tails of the distribution, our model generates less extreme values than can be observed in the real data (namely the maximum values observed are 698 for waiting time and 4000 for DDD, while in the simulated data, they are 292 and 1052, respectively), hence indicating that more heavy-tailed distributions might improve the model fit.
The patterns of simulated state sequences compared to the decoded ones of the SHI data are similar, which implies that the model can reproduce reasonable (disease) state trajectories for COPD patients.

\subsection{Limitations and discussion of findings}

When interpreting the model results, it is important to keep in mind that the physician consultations and DDDs considered in the analysis are not exclusively related to COPD, but comprise all healthcare interactions of a patient.
Consequently, the (disease) states identified correspond to general health conditions of COPD patients, meaning that a higher disease level (i.e.\ state 1) does not necessarily correspond to a worsening of COPD but could relate to any (temporary) poor health condition.
This allows us to capture a wide range of COPD-related health issues. 

As the state-dependent distributions of the mark process overlap substantially (except for the probabilities of zeros), state decoding is mainly driven by the waiting times and less by the DDDs. 
It is thus unclear if the states merely distinguish periods of high and low healthcare interactions or if they provide further information on the disease process by accounting for drug use.
In particular, the informative value of DDDs for the disease severity is questionable, given that some vitamin supplements can have much higher DDDs than COPD-specific drugs. 
However, information on specific clinical measurements like biomarkers is unavailable, since claims data are collected for billing purposes and not originally intended for scientific research. 
Yet, other variables contained in claims data --- such as costs associated with each healthcare interaction or specific treatments --- could (additionally) be used in the mark process to facilitate inference on the underlying disease levels.
Furthermore, to validate the interpretation of the states as (distinct) disease levels, the decoded state sequences of the fitted MMMPP would need to be compared to additional data on persons' true health status over time.

Regarding the estimated covariate effects in our model, the result that persons with mild ACCI spend less time in state 1 compared to those with moderate ACCI reinforces the interpretation of state 1 as a higher disease level than state 2.
Of all considered groups, male patients are expected to remain in state 2 longest, which is characterised by less frequent healthcare interactions, before switching to state 1 --- whether this is caused by different underlying disease dynamics or by the fact that men tend to wait longer before seeking medical advice than women \citep[e.g.][]{hohn2020} cannot be answered by our model. 
Furthermore, it may appear puzzling that the expected percentage of time in the high disease state considerably decreases as a function of age at diagnosis (while accounting for patients' sex and ACCI).
This effect, however, is likely confounded by our use of the age-adjusted comorbidity index (ACCI): since the ACCI adds one additional point for each decade after fifty, persons in their sixties, for instance, are only allowed two additional comorbidities to remain in the study population (as we excluded everyone with an ACCI larger than 4), whereas persons younger than fifty can have up to four additional comorbidities.
As persons with more comorbidities have a higher healthcare utilisation \citep[e.g.][]{simon2011, hutchinson2015}, the age-specific difference in (known) comorbidities thus leads to younger patients spending more time in state~1.

\section{Discussion}
\label{s:discussion}

When working with longitudinal data to understand the evolution of disease processes, a high-level decision to be made is whether observation times are (assumed to be) informative (i.e.\ dependent on the measure of interest, such as the disease severity) or non-informative. 
In particular, neglecting an informative observation process in the analysis of disease dynamics or outcomes potentially leads to biased parameter estimates \citep{gruger1991, pullenayegum2016, gasparini2020}. 
This risk of bias, however, appears to be lacking in awareness, since healthcare longitudinal studies rarely report on the potential informativeness of observation times \citep{farzanfar2017}. 
While joint models incorporating both informative observation times and disease processes exist \citep{chen2010, sweeting2010, chen2011, chen2013}, they mostly rely on data with directly observed disease stages and assume pre-scheduled examinations with informative missingness instead of patient-initiated visit times. 
Specifically for electronic health records and medical claims data, however, observation times are likely correlated with patients' latent disease dynamics: 
for example, patients with more severe conditions or acute symptoms possibly visit their physicians more often than those with mild conditions or no symptoms \citep[see, e.g.,][]{lange2015, alaa2017, gasparini2020, su2021}. 
In addition to the observation times, observed variables like patients' drug consumption or healthcare costs depend on the disease status, which is unobserved in claims data.
Therefore, we propose to infer disease dynamics from the latter using Markov-modulated marked Poisson processes.
Instead of regarding the informative observation process as a nuisance that needs to be accounted for, MMMPPs explicitly use it to infer latent (disease) states by integrating both the event times (i.e.\ the observation times of healthcare interactions) and event-specific information (i.e.\ the observed marks) into a joint model. 

Due to their flexible model structure, MMMPPs offer manifold possibilities to analyse claims data; they not only operate in continuous time and allow for (potentially multivariate) observations consisting of various data types but can also include covariate effects on the disease dynamics, the event rates, or the mark distributions. 
In many applications, however, it is not a priori clear which parameters, i.e.\ which model components, depend on covariates. 
For example, persons' sex might directly affect their event rate --- e.g.\ women might have more frequent interactions with the healthcare system than men --- or rather the disease dynamics, which then result in varying event rates --- e.g.\ women might spend more time in poor health states than men, resulting in more frequent healthcare interactions.
While model selection methods can be applied to choose between different candidate models, this task already becomes impractical for a moderate number of covariates, given that each can possibly affect all three processes. 
In addition, model selection is not only a challenge regarding covariates but also regarding the number of (disease) states underlying the observation sequences. 
As medical claims are complex data sets characterised by, for example, outliers, multimodality, and individual heterogeneity, additional states can capture otherwise neglected structure in the data and thus, significantly improve the model likelihood, which is why information criteria such as AIC and BIC usually point to models with undesirably high numbers of states --- a problem well known for HMMs \citep{pohle2017}. 
Therefore, we recommend to be pragmatic about both including covariates and selecting the number of (disease) states by taking into account expert knowledge, the current state of research, computational considerations, and especially the specific research question(s) tackled.  
To include a covariate in only one of the possible processes and to select a small, reasonable number of distinct states can help retain the model's interpretability.

Although the MMMPPs presented incorporate significant characteristics of claims data such as the informative observation process, they do not yet allow for unobserved individual heterogeneity or a flexible distribution of the duration in the states. 
While both issues are particularly relevant in the medical context \citep{cook2014}, including random effects and/or (continuous-time) semi-Markov processes in the model proves challenging.
For numerically maximising the likelihood, integrating over all possible values of random effects renders the likelihood calculation intractable, which is why other estimation techniques such as Bayesian methods \citep[e.g.][]{hui2008} or the Monte-Carlo EM algorithm \citep[e.g.][]{altman2007, chen2013} are required.
Furthermore, (continuous-time) semi-Markov processes are generally more demanding to apply than models assuming the Markov property \citep[see, e.g.,][]{chen2004, alaa2017, alaa2018}, while incorporating covariates is not straightforward anymore \citep[e.g.][]{hubbard2016}. 
Despite this potential for model extensions, MMMPPs are able to detect distinct patterns of healthcare utilisation related to disease dynamics and their (possible) association with person characteristics, as illustrated in the case study on statutory health insurance data from Germany.
In conclusion, the continuous-time latent-state approach of MMMPPs offers a natural framework to analyse the evolution of patients’ disease activity underlying claims data by jointly modelling observations and their informative time points.

\section*{Acknowledgements}

% We thank the BARMER for providing us access to their scientific data warehouse, on which all analyses in this paper were performed.
We thank Roland Langrock for stimulating discussions and helpful comments 
as well as Rebecca Louise Hilder for valuable feedback on a first draft of the manuscript. 
This research was funded by the German Research Foundation (DFG) as part of the SFB TRR 212 (NC$^3$) – Projektnummer 316099922.

\newpage

\bibliographystyle{apalike}
\bibliography{refs}

\begin{thebibliography}{}

\bibitem[Akmatov et~al., 2019]{akmatov2019}
Akmatov, M.~K., Steffen, A., Holstiege, J., and B{\"a}tzing, J. (2019).
\newblock Chronic obstructive pulmonary disease {(COPD)} in ambulatory care in
  {Germany}--temporal trends and small-area variations.
\newblock {\em Central Research Institute for Ambulatory Health Care in Germany
  (Zi). Versorgungsatlas Report}, 19/06.

\bibitem[Alaa et~al., 2017]{alaa2017}
Alaa, A.~M., Hu, S., and {van der Schaar}, M. (2017).
\newblock Learning from clinical judgments: semi-{Markov}-modulated marked
  {Hawkes} processes for risk prognosis.
\newblock In Precup, D. and Teh, Y.~W., editors, {\em Proceedings of the 34th
  International Conference on Machine Learning}, volume~70, pages 60--69. PMLR.

\bibitem[Alaa and Van Der~Schaar, 2018]{alaa2018}
Alaa, A.~M. and Van Der~Schaar, M. (2018).
\newblock A hidden absorbing semi-{Marko}v model for informatively censored
  temporal data: learning and inference.
\newblock {\em The Journal of Machine Learning Research}, 19(1):108--169.

\bibitem[Altman, 2007]{altman2007}
Altman, R.~M. (2007).
\newblock Mixed hidden {Markov} models: an extension of the hidden {Markov}
  model to the longitudinal data setting.
\newblock {\em Journal of the American Statistical Association},
  102(477):201--210.

\bibitem[Altman and Petkau, 2005]{altman2005}
Altman, R.~M. and Petkau, A.~J. (2005).
\newblock Application of hidden {Markov} models to multiple sclerosis lesion
  count data.
\newblock {\em Statistics in Medicine}, 24(15):2335--2344.

\bibitem[Amoros et~al., 2019]{amoros2019}
Amoros, R., King, R., Toyoda, H., Kumada, T., Johnson, P.~J., and Bird, T.~G.
  (2019).
\newblock A continuous-time hidden {Markov} model for cancer surveillance using
  serum biomarkers with application to hepatocellular carcinoma.
\newblock {\em Metron}, 77:67--86.

\bibitem[Bannay et~al., 2016]{bannay2016}
Bannay, A., Chaignot, C., Bloti{\`e}re, P.-O., Basson, M., Weill, A.,
  Ricordeau, P., and Alla, F. (2016).
\newblock The best use of the {Charlson} comorbidity index with electronic
  health care database to predict mortality.
\newblock {\em Medical Care}, 54(2):188--194.

\bibitem[Bureau et~al., 2003]{bureau2003}
Bureau, A., Shiboski, S., and Hughes, J.~P. (2003).
\newblock Applications of continuous time hidden {Markov} models to the study
  of misclassified disease outcomes.
\newblock {\em Statistics in Medicine}, 22(3):441--462.

\bibitem[Chen et~al., 2010]{chen2010}
Chen, B., Yi, G.~Y., and Cook, R.~J. (2010).
\newblock Analysis of interval-censored disease progression data via
  multi-state models under a nonignorable inspection process.
\newblock {\em Statistics in Medicine}, 29(11):1175--1189.

\bibitem[Chen and Zhou, 2011]{chen2011}
Chen, B. and Zhou, X.-H. (2011).
\newblock Non-homogeneous {Markov} process models with informative observations
  with an application to {Alzheimer's} disease.
\newblock {\em Biometrical Journal}, 53(3):444--463.

\bibitem[Chen and Zhou, 2013]{chen2013}
Chen, B. and Zhou, X.-H. (2013).
\newblock A correlated random effects model for non-homogeneous {Markov}
  processes with nonignorable missingness.
\newblock {\em Journal of Multivariate Analysis}, 117:1--13.

\bibitem[Chen and Tien, 2004]{chen2004}
Chen, P.-L. and Tien, H.-C. (2004).
\newblock Semi-{Markov} models for multistate data analysis with periodic
  observations.
\newblock {\em Communications in Statistics - Theory and Methods},
  33(3):475--486.

\bibitem[Choquet, 2018]{choquet2018}
Choquet, R. (2018).
\newblock Markov-modulated {P}oisson processes as a new framework for analysing
  capture-recapture data.
\newblock {\em Methods in Ecology and Evolution}, 9(4):929--935.

\bibitem[Christensen et~al., 2018]{christensen2018}
Christensen, T., Frandsen, A., Glazier, S., Humpherys, J., and Kartchner, D.
  (2018).
\newblock Machine learning methods for disease prediction with claims data.
\newblock In {\em 2018 IEEE International Conference on Healthcare Informatics
  (ICHI)}, pages 467--4674.

\bibitem[Cook and Lawless, 2014]{cook2014}
Cook, R.~J. and Lawless, J.~F. (2014).
\newblock Statistical issues in modeling chronic disease in cohort studies.
\newblock {\em Statistics in Biosciences}, 6(1):127--161.

\bibitem[Dutta et~al., 2022]{dutta2022}
Dutta, E.~K., Kumar, S., Venkatachalam, S., Downey, L.~E., and Albert, S.
  (2022).
\newblock An analysis of government-sponsored health insurance enrolment and
  claims data from {Meghalaya}: Insights into the provision of health care in
  {North East India}.
\newblock {\em PloS One}, 17(6):e0268858.

\bibitem[Farzanfar et~al., 2017]{farzanfar2017}
Farzanfar, D., Abumuamar, A., Kim, J., Sirotich, E., Wang, Y., and
  Pullenayegum, E. (2017).
\newblock Longitudinal studies that use data collected as part of usual care
  risk reporting biased results: a systematic review.
\newblock {\em BMC Medical Research Methodology}, 17:133.

\bibitem[Forney, 1973]{forney1973}
Forney, G.~D. (1973).
\newblock The {Viterbi} algorithm.
\newblock {\em Proceedings of the IEEE}, 61(3):268--278.

\bibitem[Fouchet et~al., 2016]{fouchet2016}
Fouchet, D., Santin-Janin, H., Sauvage, F., Yoccoz, N.~G., and Pontier, D.
  (2016).
\newblock An {R} package for analysing survival using continuous-time open
  capture-recapture models.
\newblock {\em Methods in Ecology and Evolution}, 7(5):518--528.

\bibitem[Gasparini et~al., 2020]{gasparini2020}
Gasparini, A., Abrams, K.~R., Barrett, J.~K., Major, R.~W., Sweeting, M.~J.,
  Brunskill, N.~J., and Crowther, M.~J. (2020).
\newblock Mixed-effects models for health care longitudinal data with an
  informative visiting process: A {Monte Carlo} simulation study.
\newblock {\em Statistica Neerlandica}, 74(1):5--23.

\bibitem[Gruger et~al., 1991]{gruger1991}
Gruger, J., Kay, R., and Schumacher, M. (1991).
\newblock The validity of inferences based on incomplete observations in
  disease state models.
\newblock {\em Biometrics}, 47(2):595--605.

\bibitem[H{\"o}hn et~al., 2020]{hohn2020}
H{\"o}hn, A., Gampe, J., Lindahl-Jacobsen, R., Christensen, K., and Oksuyzan,
  A. (2020).
\newblock Do men avoid seeking medical advice? {A} register-based analysis of
  gender-specific changes in primary healthcare use after first hospitalisation
  at ages 60+ in {Denmark}.
\newblock {\em Journal of Epidemiology and Community Health}, 74(7):573--579.

\bibitem[Hossain et~al., 2019]{hossain2019}
Hossain, M.~E., Khan, A., Moni, M.~A., and Uddin, S. (2019).
\newblock Use of electronic health data for disease prediction: A comprehensive
  literature review.
\newblock {\em IEEE/ACM Transactions on Computational Biology and
  Bioinformatics}, 18(2):745--758.

\bibitem[Hubbard et~al., 2016]{hubbard2016}
Hubbard, R., Lange, J., Zhang, Y., Salim, B., Stroud, J., and Inoue, L. (2016).
\newblock Using semi-{Markov} processes to study timeliness and tests used in
  the diagnostic evaluation of suspected breast cancer.
\newblock {\em Statistics in Medicine}, 35(27):4980--4993.

\bibitem[Hui-Min~Wu et~al., 2008]{hui2008}
Hui-Min~Wu, G., Chang, S.-H., and Hsiu-Hsi~Chen, T. (2008).
\newblock A {Bayesian} random-effects {Markov} model for tumor progression in
  women with a family history of breast cancer.
\newblock {\em Biometrics}, 64(4):1231--1237.

\bibitem[Hutchinson et~al., 2015]{hutchinson2015}
Hutchinson, A.~F., Graco, M., Rasekaba, T.~M., Parikh, S., Berlowitz, D.~J.,
  and Lim, W.~K. (2015).
\newblock Relationship between health-related quality of life, comorbidities
  and acute health care utilisation, in adults with chronic conditions.
\newblock {\em Health and Quality of Life Outcomes}, 13:69.

\bibitem[Jackson et~al., 2003]{jackson2003}
Jackson, C.~H., Sharples, L.~D., Thompson, S.~G., Duffy, S.~W., and Couto, E.
  (2003).
\newblock Multistate {Markov} models for disease progression with
  classification error.
\newblock {\em The Statistician}, 52(2):193--209.

\bibitem[Lange et~al., 2015]{lange2015}
Lange, J.~M., Hubbard, R.~A., Inoue, L. Y.~T., and Minin, V.~N. (2015).
\newblock A joint model for multistate disease processes and random informative
  observation times, with applications to electronic medical records data.
\newblock {\em Biometrics}, 71(1):90--101.

\bibitem[Langrock et~al., 2013]{langrock2013}
Langrock, R., Borchers, D.~L., and Skaug, H.~J. (2013).
\newblock Markov-modulated nonhomogeneous {Poisson} processes for modeling
  detections in surveys of marine mammal abundance.
\newblock {\em Journal of the American Statistical Association},
  108(503):840--851.

\bibitem[Lu, 2012]{lu2012}
Lu, S. (2012).
\newblock Markov modulated {Poisson} process associated with state-dependent
  marks and its applications to the deep earthquakes.
\newblock {\em Annals of the Institute of Statistical Mathematics},
  64(1):87--106.

\bibitem[Luo et~al., 2021]{luo2021}
Luo, Y., Stephens, D.~A., Verma, A., and Buckeridge, D.~L. (2021).
\newblock Bayesian latent multi-state modeling for nonequidistant longitudinal
  electronic health records.
\newblock {\em Biometrics}, 77(1):78--90.

\bibitem[Mews et~al., 2022]{mews2022}
Mews, S., Langrock, R., King, R., and Quick, N. (2022).
\newblock Multi-state capture-recapture models for irregularly sampled data.
\newblock {\em The Annals of Applied Statistics}, 16(2):982--998.

\bibitem[Min et~al., 2019]{min2019}
Min, X., Yu, B., and Wang, F. (2019).
\newblock Predictive modeling of the hospital readmission risk from patients'
  claims data using machine learning: a case study on {COPD}.
\newblock {\em Scientific Reports}, 9:2362.

\bibitem[Moler and Van~Loan, 2003]{moler2003}
Moler, C. and Van~Loan, C. (2003).
\newblock Nineteen dubious ways to compute the exponential of a matrix,
  twenty-five years later.
\newblock {\em SIAM Review}, 45(1):3--49.

\bibitem[Nerius et~al., 2017]{nerius2017}
Nerius, M., Fink, A., and Doblhammer, G. (2017).
\newblock Parkinson's disease in {Germany}: prevalence and incidence based on
  health claims data.
\newblock {\em Acta Neurologica Scandinavica}, 136(5):386--392.

\bibitem[Nielsen et~al., 2017]{nielsen2017}
Nielsen, S.~S., Warden, M.~N., Camacho-Soto, A., Willis, A.~W., Wright, B.~A.,
  and Racette, B.~A. (2017).
\newblock A predictive model to identify {Parkinson} disease from
  administrative claims data.
\newblock {\em Neurology}, 89(14):1448--1456.

\bibitem[Ploner et~al., 2020]{ploner2020}
Ploner, T., He{\ss}, S., Grum, M., Drewe-Boss, P., and Walker, J. (2020).
\newblock Using gradient boosting with stability selection on health insurance
  claims data to identify disease trajectories in chronic obstructive pulmonary
  disease.
\newblock {\em Statistical Methods in Medical Research}, 29(12):3684--3694.

\bibitem[Pohle et~al., 2017]{pohle2017}
Pohle, J., Langrock, R., van Beest, F.~M., and Schmidt, N.~M. (2017).
\newblock Selecting the number of states in hidden {Markov} models: pragmatic
  solutions illustrated using animal movement.
\newblock {\em Journal of Agricultural, Biological and Environmental
  Statistics}, 22(3):270--293.

\bibitem[Pullenayegum and Lim, 2016]{pullenayegum2016}
Pullenayegum, E.~M. and Lim, L.~S. (2016).
\newblock Longitudinal data subject to irregular observation: A review of
  methods with a focus on visit processes, assumptions, and study design.
\newblock {\em Statistical Methods in Medical Research}, 25(6):2992--3014.

\bibitem[Quan et~al., 2011]{quan2011}
Quan, H., Li, B., Couris, C.~M., Fushimi, K., Graham, P., Hider, P., Januel,
  J.-M., and Sundararajan, V. (2011).
\newblock Updating and validating the {Charlson} comorbidity index and score
  for risk adjustment in hospital discharge abstracts using data from 6
  countries.
\newblock {\em American Journal of Epidemiology}, 173(6):676--682.

\bibitem[Scott and Smyth, 2003]{scott2003}
Scott, S.~L. and Smyth, P. (2003).
\newblock The {Markov} modulated {Poisson} process and {Markov Poisson} cascade
  with applications to web traffic modeling.
\newblock In Bernardo, J.~M., Bayarri, M.~J., Berger, J.~O., Dawid, A.~P.,
  Heckerman, D., Smith, A. F.~M., and West, M., editors, {\em Bayesian
  Statistics 7}, pages 671--680. Oxford University Press, Oxford.

\bibitem[Simon-Tuval et~al., 2011]{simon2011}
Simon-Tuval, T., Scharf, S.~M., Maimon, N., Bernhard-Scharf, B.~J., Reuveni,
  H., and Tarasiuk, A. (2011).
\newblock Determinants of elevated healthcare utilization in patients with
  {COPD}.
\newblock {\em Respiratory Research}, 12:7.

\bibitem[Su et~al., 2021]{su2021}
Su, L., Cheng, Y., Pereira, D.~I., and Powell, J.~J. (2021).
\newblock Modelling disease progression with multi-level electronic health
  records data and informative observation times: an application to treating
  iron deficiency anaemia in primary care of the {UK}.
\newblock {\em arXiv preprint arXiv:2107.13956}.

\bibitem[Sweeting et~al., 2010]{sweeting2010}
Sweeting, M., Farewell, V., and De~Angelis, D. (2010).
\newblock Multi-state {Markov} models for disease progression in the presence
  of informative examination times: An application to hepatitis {C}.
\newblock {\em Statistics in Medicine}, 29(11):1161--1174.

\bibitem[Viterbi, 1967]{viterbi1967}
Viterbi, A. (1967).
\newblock Error bounds for convolutional codes and an asymptotically optimum
  decoding algorithm.
\newblock {\em IEEE Transactions on Information Theory}, 13(2):260--269.

\bibitem[Zucchini et~al., 2016]{zucchini2016}
Zucchini, W., MacDonald, I.~L., and Langrock, R. (2016).
\newblock {\em Hidden {Markov Models} for {Time Series}: {An Introduction}
  {U}sing {R}}.
\newblock Chapman \& Hall/CRC, Boca Raton.

\end{thebibliography}

% \newpage

\appendix

\counterwithin{figure}{section}
\counterwithin{table}{section}

\section{Appendix}
\label{app1}

\subsection{Selection criteria for the case study data}

From the statutory health insurance (SHI) data set, we selected persons with a confirmed in- or outpatient diagnosis of COPD (based on the ICD-10 code J44) in the first and second quarter of 2008 who had no prior COPD diagnosis within a three-year pre-observation period (2005--2007).
To ensure patients' chronic condition and to avoid any misdiagnoses in the data, we further required a permanent treatment within the year following initial COPD diagnosis (i.e.\ at least one more COPD diagnosis in each of the following three quarters). 
Persons who were younger than 18, who died during the study period, or who were observed for less than six years after their initial COPD diagnosis were excluded from the analysis.
Furthermore, to reduce heterogeneity in our study population, we considered only persons whose overall costs during the study period was less than the 95\% quantile of costs from all COPD patients with initial diagnosis in the years 2008--2015.
As we cannot rule out the possibility of missing observations in the SHI data, we removed persons without any healthcare interactions over a period of more than two years from the study population. 
Lastly, persons with less than 40 observations after their initial COPD diagnosis (corresponding to less than 1\% of the data) were excluded from the analysis, as the focus here lies on disease progression through different stages, which can only be observed based on a sufficient number of observations.

\newpage

\subsection{Additional information on the results of modelling disease activity for COPD patients}

\begin{figure}[h]
    \centering 
    \includegraphics[width = 0.85\textwidth]{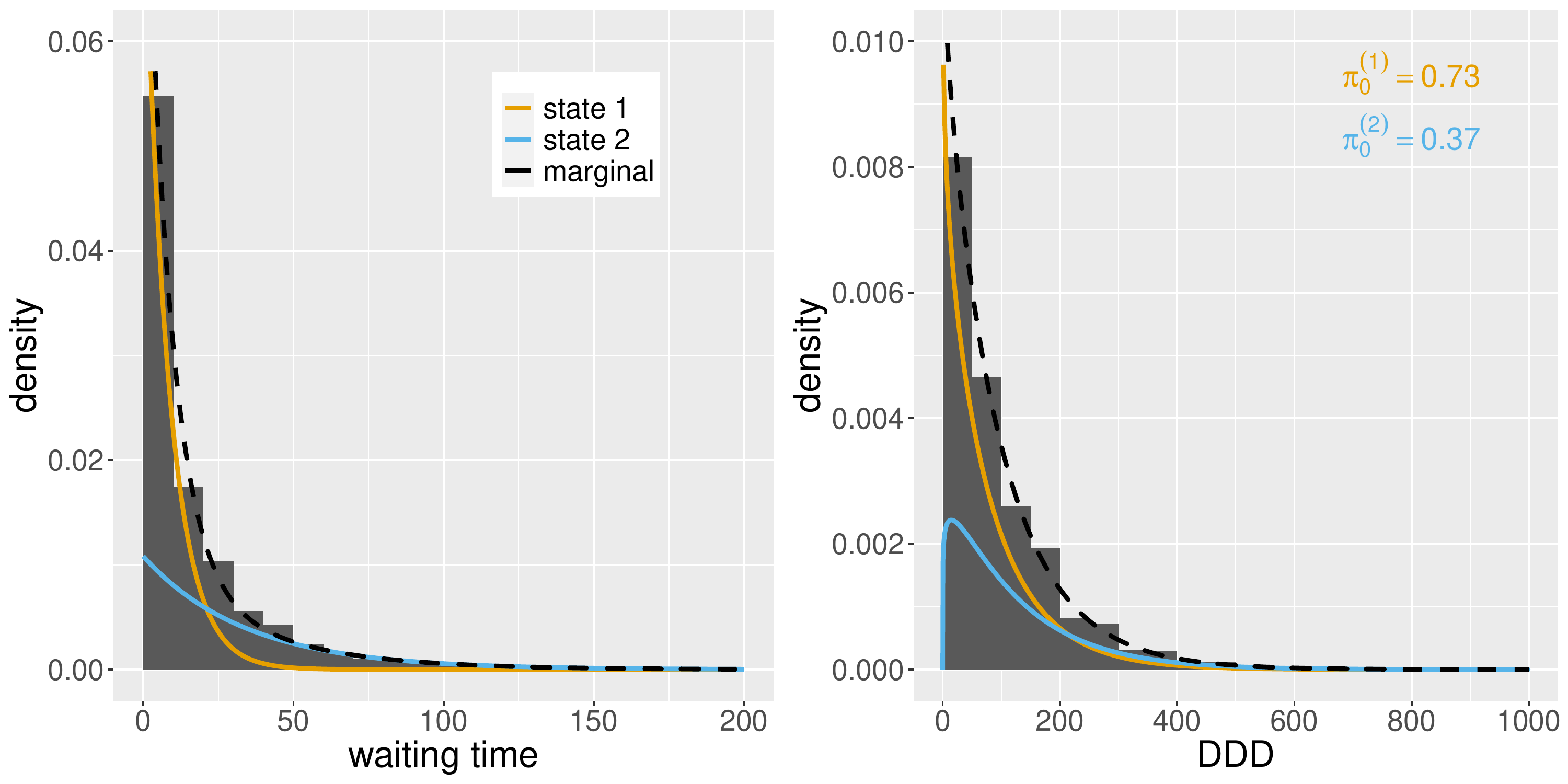}
    \caption{Estimated state-dependent distributions for the observation process (exponential distributions; left) and the mark process (zero-adjusted gamma distributions; right) with histograms of the observed waiting times and DDDs, respectively. The state-dependent distributions are weighted by the percentage of time spent in the different states, as calculated based on the decoded state sequences, while the dashed lines are the associated marginal distributions.}
    \label{fig:histDens}
    \vspace*{-15mm}
\end{figure}

\vspace*{20mm}

\begin{table}[hbt]
    \centering
    \caption{Estimated regression coefficients on the state transition intensities $q_{12}$ and $q_{21}$, respectively, with 95\% confidence intervals (CIs). The latter were calculated based on the observed Fisher information.} % obtained in the case study 
    \label{tab:coefSI}
    \begin{tabular}{l c c c c} 
    \\[-1.8ex]\hline 
\hline \\[-1.8ex] 
         & $\beta_0^{12}$ & $\beta_1^{12}$ & $\beta_2^{12}$ & $\beta_3^{12}$  \\ \hline
         estimate & -3.87 & -0.02 & -0.11 & 0.08 \\ \vspace{3mm}
         95\% CI & [-3.94; -3.81] & [-0.12; 0.08] & [-0.23; 0.02] & [0.03; 0.13] \\ \hline
         & $\beta_0^{21}$ & $\beta_1^{21}$ & $\beta_2^{21}$ & $\beta_3^{21}$ \\ \hline
         estimate & -4.55 & -0.62 & -0.45 & -0.07 \\ \vspace{3mm}
         95\% CI & [-4.61; -4.48] & [-0.72; -0.52] & [-0.58; -0.33] & [-0.12; -0.02] \\
         \hline
    \end{tabular}
\end{table}

\begin{figure}[htb]
    \centering 
    \includegraphics[width = 0.8\textwidth]{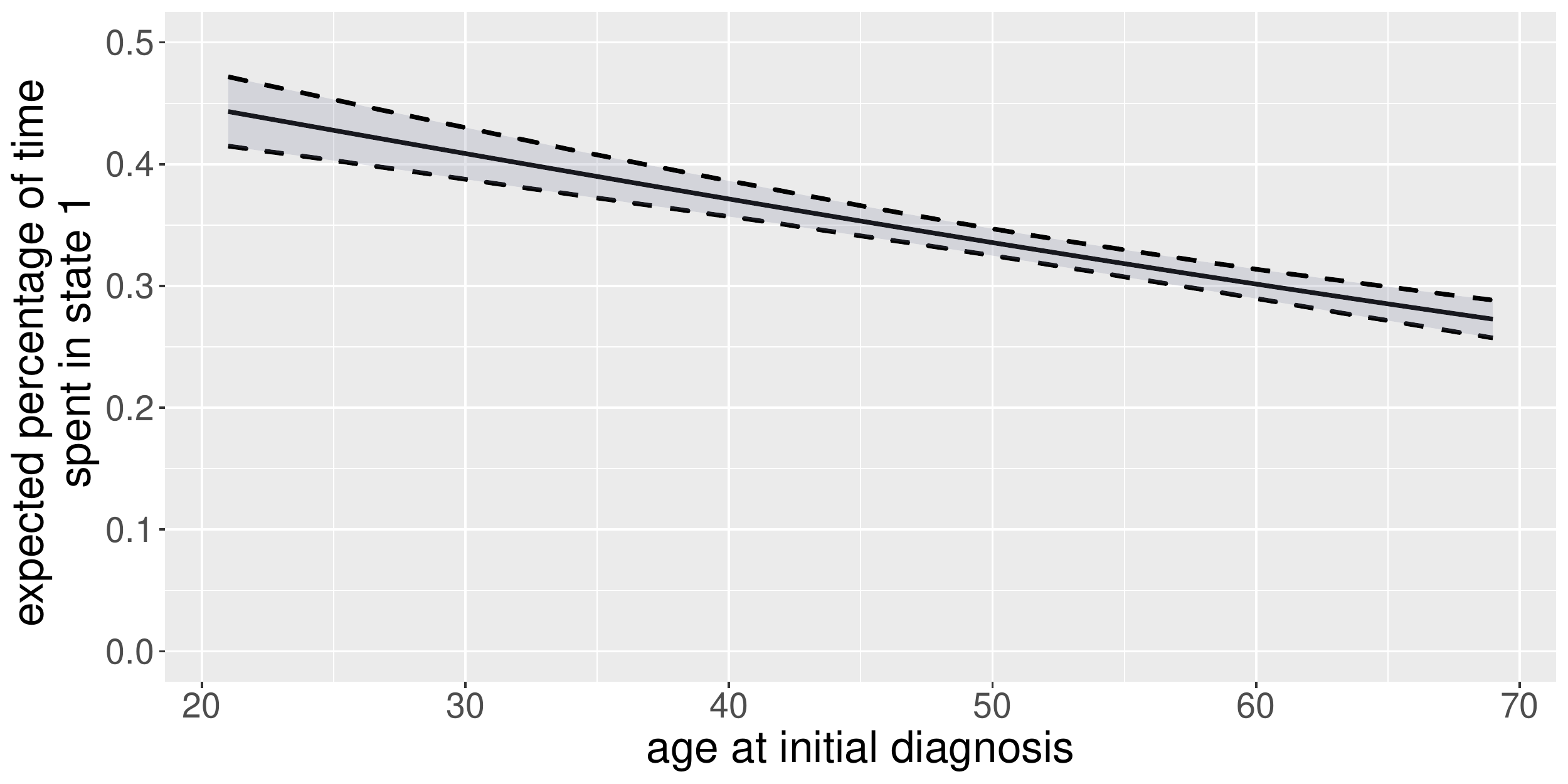}
    \caption{Expected percentage of time spent in state 1 (i.e.\ the high disease level; mean and 95\% CI) under the fitted model as a function of the covariate age at initial diagnosis, given that all other covariates are fixed to their reference values, and plotted for the range of observed values in the study population (i.e.\ ages 21--69). CIs were obtained based on Monte Carlo simulation from the estimators’ approximate distribution as implied by maximum likelihood theory. The estimated coefficients underlying this figure are provided in Table~\ref{tab:coefSI} in the Appendix.}
    \label{fig:ageEffect}
\end{figure}

\begin{figure}[hb]
    \centering
    \includegraphics[width = 0.85\textwidth]{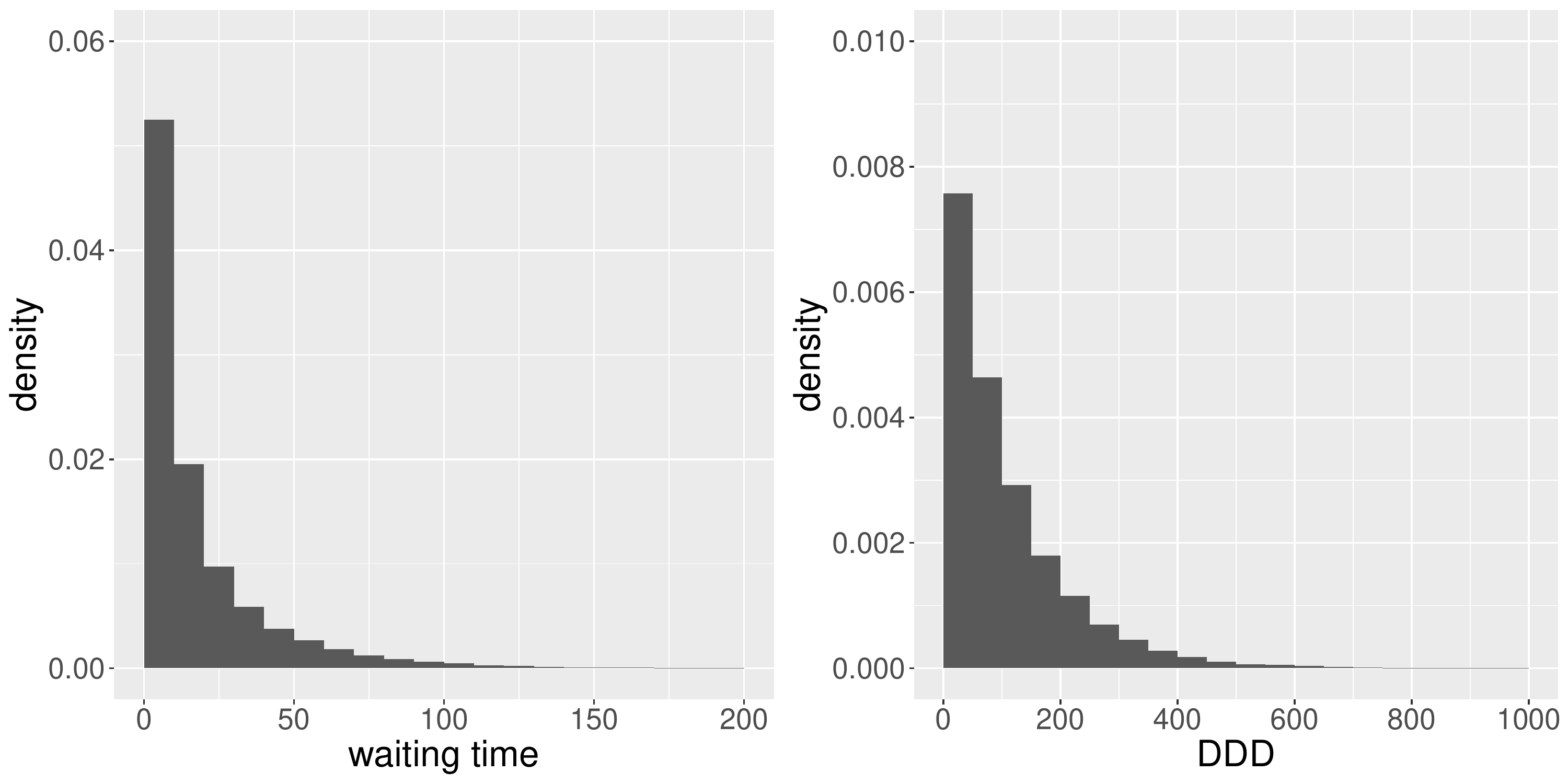}
    \caption{Histograms of the simulated waiting times (left) and DDDs (right).}
    \label{fig:simDist}
    %\vspace*{5in}
\end{figure}

\end{spacing}

\end{document}